# Ribosome collisions & Translation efficiency: Optimization by codon usage and mRNA destabilization.

Namiko Mitarai[1, 3] Kim Sneppen[1] and Steen Pedersen[2*]

[1] Center for Models of Life, Niels Bohr Institute, and [2] Department of Biology, University of Copenhagen, Denmark.

[3] Department of Physics, Kyushu University, Japan.

* Corresponding author: Steen Pedersen, Department of Biology, University of Copenhagen,

Ole Maaløes vej 5, DK-2200 Copenhagen N, Denmark

Electronic Mail: steenp@bio.ku.dk

Telephone (+45) 3532 2096, Fax: (+45) 3532 2128

**Abstract**

Individual mRNAs are translated by multiple ribosomes that initiate translation with a few seconds interval. The ribosome speed is codon dependant, and ribosome queuing has been suggested to explain specific data for translation of some mRNAs *in vivo*. By modelling the stochastic translation process as a traffic problem, we here analyze conditions and consequences of collisions and queuing. The model allowed us to determine the on-rate (0.8 to 1.1 initiations per sec) and the time (1 sec) the preceding ribosome occludes initiation for *Escherichia coli lacZ* mRNA *in vivo*. We find that ribosome collisions and queues are inevitable consequences of a stochastic translation mechanism that reduce the translation efficiency substantially on natural mRNAs. The cells minimize collisions by having its mRNAs being unstable and by a highly selected codon usage in the start of the mRNA. The cost of mRNA breakdown is offset by the concomitant increase in translational efficiency.

**Introduction**

The degeneracy of the genetic code opens for additional layers of coding information of importance for protein expression and regulation in response to a changing environment. One of these layers is associated to the fact that different codons are translated with widely different rates, thereby providing regulatory flexibility at the translational level (1). Furthermore, it has been found that subsequent ribosomes are loaded onto the mRNA sufficiently fast to make them interfere with each other, and sometimes even form extended queues causing a substantial delay for producing proteins from older mRNAs (2). As a consequence the translation process itself influences the cellular expression of proteins. To quantify the importance of codon to codon variation in translation rate one need to develop a mathematical model and to calibrate it to available data on mRNA translation kinetics.

Previously, we developed a method to measure the time it takes for ribosomes to synthesize peptides in living cells on individual mRNA species (1) and refined it to have a standard error, SEM of about 0,5 sec when measuring translation times of mRNA encoding abundant proteins (3). We found that mRNAs with the ribosomal protein codon usage (4) were translated approximately 35 % faster than mRNAs with a more unbiased codon usage as for instance the *lacI* mRNA (1). This rate difference was determined by the codon usage and not influenced significantly by potential hairpin structures in the slowly translated mRNAs (3). A subsequent study showed that two synonymous codons read by the same tRNA species were translated with a three-fold difference in rate, implying that the slow translation of these codons were not due to limits of tRNA abundance. Furthermore, inserting the sequence $CGA_3(GAG)_{24}CCG_3$ two times in the *lacZ* mRNA provoked queue formation[1] and reduced the expression slightly, probably because of transcriptional polarity, whereas the same sequence inserted only once gave no detectable ribosome queues and no reduced expression (2).

To analyze the interplay between the stochastic movement of ribosomes and queue formation we determine key parameters associated to translation of the *lacZ* mRNA by introducing a model that reproduces the previous experiments (2). The model is an extension of the stochastic hopping processes, considered for uniform translation rates (5, 6) and extended to codon-dependent translation rates (7,8, 9). Also, Ringquist et al. (10) investigated a number of variables that determined the strength of the *lacZ* ribosome-binding site, RBS[2], including the strength of the SD interaction, the distance between the SD and the initiation codon and the nature of the initiation codon. They concluded that their data were consistent with a simple kinetic model in which a variety of rate constants contribute to the process of translation initiation. Our work attempts to specify these rates by reproducing experimental data on *lacZ* translation with and without slow inserts in a model. We now can analyse how codon usage and mRNA stability may be used to influence speed, expression yield and metabolic cost associated to protein production.

---

[1] We define collision to mean the physical hindrance to translation by a preceding ribosome and queuing to mean extensive collisions where more than two ribosomes are involved.

[2] Abbreviations RBS: ribosome-binding site; SD: the Shine-Dalgarno sequence complementary to the 3' end of 16S rRNA; CAI codon adaptation index.

## Results

### Description of the model of translation kinetics

First we model the basic dynamics of translation. Fig. 1 depicts ribosomes that translate an mRNA. Translation starts by binding a 30S ribosomal subunit to codon 1 with the rate $K_s$, provided that the binding site is accessible, which requires that no ribosome is within the occluding distance, $d$ from the binding site. The ribosome stays at codon 1 for a time, $\tau$, to assemble the translating 70S ribosome. Then, for each time-step $dt$ after $\tau$, all ribosomes have a probability $R_x dt$ to move one codon forward, where $R_x$ is the translation rate for the codon at position $x$. Possible movements are accepted only if the distance to the preceding ribosome is larger than $d$. When the ribosome reaches the end of the mRNA, the peptide is released with a rate $K_t$.

We also model the finite life-time of the mRNA by simulating an ensemble of mRNAs produced at random times and subsequently degraded with the rate that reproduces an average of $p$ translations per mRNA. When mRNA degradation is initiated, translation initiation is stopped and already initiated ribosomes continue translation until terminating at the 3' end as suggested by previous experiments (11, 12).

The detailed description of the simulation procedure is given in the Methods section. The java applet of the model can be found at URL: http://cmol.nbi.dk/models/RibosomeTraffic/RibosomeTraffic.html

### Description of the experiment

Using this model of translation dynamics, we reproduce the *in vivo* experiment for incorporation of radioactivity in a model protein (2). The experiment is performed as follows: At time $t=0$ the ensemble of mRNA's being translated in steady state is exposed to a 10 seconds pulse of radioactive $^{35}$S-methionine. The pulse is stopped by adding a large excess of unlabelled methionine to the culture. After this chase, samples are taken at 10 sec intervals and the radioactivity in the completed protein determined. Fig. 2 shows a schematic incorporation curve. The sketch of the mRNA along the ordinate axis illustrates that shortly after the $^{35}$S-Met pulse and chase, only ribosomes that were close to terminating at the time of the pulse will contribute radioactivity to the finished peptide. As time progresses and more samples are taken, more and more radioactivity will accumulate in the finished peptide. When the first ribosome on the mRNA at the time of the pulse has terminated translation, the incorporation of radioactivity will stop. This defines the translation time as the time it takes to bring the most N terminal methionine into the full-length protein.

To minimize differences among the experiments a mixture of two cultures was used, one with the parental plasmid pMAS23, and one encoding *lacZ* with the insert to be compared (2). The direct comparison in the same experiment was possible because the two full-length proteins were resolved by gel electrophoresis. Furthermore, a double isotope labelling technique enabled correction for a variable recovery during sample preparation. Normally, 18 data points were collected in an experiment. From these incorporation curves and the position of methionine codons in the mRNA, indicated in Fig 1, the average translation time of these mRNAs were determined with a SEM of approximately 0.5 sec. Comparing mRNA, with and without an insert, the time for translating the insert was determined (2).

To reproduce this experiment in our model analysis, we prepare the steady state ensemble of mRNAs, count the incorporation of the radioactive methionine for the ribosomes that pass each methionine codon during the pulse and calculate how many radioactive methionine residues are included in the finished peptide at each time. This gives the incorporation curve from the simulation and it is compared with the experimental incorporation curves. Further aspects of the experimental and model methods are discussed in Methods.

**Input parameters to the model.**

The rate Ks, of binding the 30S ribosome to the mRNA can be set in the simulation. For the resulting *lacZ* mRNA initiation has been estimated to occur about once per 3 sec (2, 11). In our analyses we test on-rates from 1 per 0.8 sec to 1 per 3 sec that will result in initiation rates from approximately 1 per 2 sec to 1 per 4 sec because it takes approximately 0.2 sec to initiate and one second to translate the first 11 codons in the *lacZ* mRNA (see later). The rate $K_t$ to release the peptide at the end of the mRNA is set to 2 per sec, which is faster than the initiation rate and does not provoke ribosome queuing. Thus, this finite termination rate does not influence the translation dynamics in the wild type situation we consider here.

The occluding distance *d* can be estimated to 11 codons because a) 30S ribosomes protects 30-35 nucleotides against RNase degradation (13); b) the 33 nucleotides from -20 to + 13 are non-random and involved in 30S ribosome binding (14) and c) 30S and 70S ribosomes occupy the same space on the mRNA (14).

An important ingredient in our model is the fact that each mRNA is only used for a limited period. The *lacZ* mRNA was estimated to be translated an average of 40 times (11) and the *trp* mRNA 20 times (15). The estimate of 40 proteins per *lacZ* mRNA is perhaps too high: As discussed by Kennell and Riezman, (11) the number for *lacZ* mRNA is critically dependent on the estimate of the enzyme specific activity of purified β-galactosidase. If the 400.000 enzyme units per mg protein (11) is replaced by the 800.000 units per mg protein found in a slightly different buffer (16), the estimated number of protein per mRNA would drop from 40 to 20. The average number for translating the *lacZ* mRNA is therefore likely to be closer to 30 and maybe even as low as 20, which is close to the overall mRNA average translation number in minimal medium (17). Thus, in our model analysis, we normally set the average translation number to 30 unless otherwise noted.

During initiation, the three initiation factors and the charged initiator tRNA may bind the 30S subunit before or after this binds to the mRNA (18) cited in (14). After forming the 30S mRNA complex the 50S subunit bind and translation can start. The time for these reactions is likely to be short, but because our method determines the time for bringing the methionine encoded by codon 3 in the gene into the completed peptide any time spent in the initiation process is not influencing the incorporation curve (see Methods). The model operates with a time $\tau$, where the ribosome is bound to the mRNA without translating to 0.2 sec, but this time might as well be included in the time between initiations.

As the last input to the simulation, we assign one of two sets of translation times to the individual codons. In the first set all codons are translated at 13.25 codons per sec that gives the observed translation time for the *lacZ* mRNA (3), see later. In the second set, we assign one of three values A, B and C for each codon specific translation rate. As explained and justified in the legend to Table 1 the A rate is approximately three fold higher then the B rate that again is approximately three fold higher than the C rate. The A, B and C rates correspond to high, medium and small CAI values for *E. coli* (19). The rates A, B, C are modelling parameters and varied in our curve fitting procedure (see later). In the discussion we analyse the robustness of the conclusions with respect to the assignment of model parameters.

The experimental data used in our modelling are shown in Fig. 3a. The incorporated radioactivity is normalized to the 23 methionine residues in the *lacZ* gene. They are obtained as the average of the data of (2) and an independent unpublished measurement. The parental plasmid pMAS23 and the mRNAs with inserts containing many GAA or GAG codons are briefly described in Methods.

The radioactivity incorporated into full-length protein increases monotonously with time in these experiments. As shown in Fig. 3a the incorporation curve for the pMAS24GAG mRNA with the short insert is displaced with a constant time from the parental pMAS23 curve early in the experiment. This interval is the time for translating the insert and the early displacement directly reflects that the insert is placed close to the 3' end of the mRNA. For the mRNA with the tandem insert the incorporation curve is progressively delayed relative to the two other curves. This more than additive effect of the long insert is simplest explained by extended queuing from the long slow insert, where ribosomes that were most 5' on the mRNA at the time of the radioactive pulse not only have to translate the insert but also have to wait for the ribosomes in front to pass the insert. This observation of a queue for the long insert in pMAS48GAG without a noticeable queue for the short insert, pMAS24GAG turned out to constrain the model parameters describing the translation process, and give a glimpse into the noise associated to the translating ribosome.

**Results from the modelling**

Fig. 3 shows several examples of the model simulation, representing the best fit to the experimental data as well as a number of variants that illustrates the most important parameters. The best fit was selected from combinations of following values of A, B and C rates: A; 20, 25, 30, and 35 codons per sec, B; 6, 7, 8, and 9 codons per sec, and C; 3, 3.5, 4, and 4.5 codons per sec.

To choose the parameter sets that best fit all the data, we calculate the variance between each simulated curve and the 13 experimental data points from the rising part of the experimental curve for pMAS23, pMAS24GAG, and pMAS48GAG. Then, we chose the parameter sets that minimize the variance for the data sets.

Fig 3B shows the modelled curves with the very best fit: the rates A, B, C being 35, 8, and 4.5 per sec, respectively, the on rate 0.9 per sec. Table 2 gives several sets of A, B and C rates, that almost equally well reproduce the experimental data. In Fig 3C, the simulation with the same parameter as Fig 3B but with deterministic translation is shown, where a ribosome waits exactly for the duration $1/R_x$ before it moves to the next codon (Details are given in Methods). In Fig.3C, all curves rise faster than Fig.3B, and the queue formation in

pMAS48GAG is not reproduced. In order to reproduce the divergence of the later part of the pMAS24GAG and pMAS48GAG curves, the stochastic translation is needed. Finally, Fig 3D demonstrates the necessity of having the A, B, C rates different. The figure shows a fully stochastic simulation with the on rate 0.9 per sec as Fig. 3B but the A, B, C rates are all 13.25 per sec. This can reproduce the data for pMAS23, but not for the mRNAs with inserts.

Finally, Fig. 4 shows a space-time plot given by the simulation. Here each line depicts the movement of a ribosome down the mRNA. Where the lines contact each other, this indicates collision between ribosomes. Fig. 4 illustrates that the short insert only rarely provoke ribosome queuing when translating mRNA an average 30 times, whereas the tandem insert regularly generate extensive queuing. The panel D show the extensive queuing arising if the pMAS48GAG mRNA had been stable. The space-time plots also showed that collisions normally are resolved quickly whereas queues once formed, often takes a long time to untangle.

The higher density of ribosomes in the beginning of the mRNA is seen in all panels. This is because the codons we have designated B and C are predominantly found here (20). We propose two arguments for the relatively high abundance of the rare and slow codons in the beginning of genes: 1) The occlusion time will respond to changes in substrate level if the rare codons, for instance GAG, is unsaturated whereas the fast translated codon GAA are saturated (GAA and GAG are read by same tRNA but are translated with A and B rates respectively). This allows negative feedback to fine-tune the initiation rate. 2) Slowly translated codons in the beginning of the mRNA will increase the average ribosome distance in the later part of the gene and thereby minimize queues there. The codon distribution found (20) can therefore be thought of as a way to increase the ribosome efficiency.

To show that the enrichment for "B and C" codons in the mRNA starts are highly selected for, we determined the frequency of synonymous codon changes in the early part of 5 highly expressed genes in species closely related to *E. coli* and compared it to the frequency of such changes in the distal part of the same genes. This analysis, Table 3, showed that the selection pressure for a certain codon usage is higher in the beginning of genes compared to the distal parts. Treating the frequencies for synonymous codon changes in each end of these genes as coupled variables in a t-test gave a less that 0.2% probability for that this difference in the frequency of synonymous codon replacements is a coincidence.

**Advantage of unstable mRNAs for translation efficiency.** As ribosomes are loaded on the mRNA with relatively short time intervals, the stochastic translation relatively fast causes ribosome-ribosome collisions. As each collision delay the lagging ribosome, there is self-perpetuating tendency to pile up of ribosomes on slow parts of the mRNA. Even when there are too few collisions to cause long queues, we find that collisions will occur, and will occur more frequently as the mRNA gets older. Our model allows us to calculate the average translation time of the mRNA as function of the average number of translations before it is degraded. Fig. 5a shows how the *lacZ* mRNA translation time is predicted to increase with the average number of translations of the mRNA. Already at 30 translations per mRNA the stochastic effects has increased the average translation time by 5% compared to an mRNA that was only translated once. Fig. 5a also shows that a stable mRNA would be translated 1.7% slower than the present *lacZ* mRNA translated on average 30 times.

This suggested to us that bacteria might have evolved unstable mRNAs as yet another way to untangle and minimize ribosome queues. We now compare the cost associated to mRNA production and recycling against the cost of making the ribosomes that would compensate for the ones that waste their time by collisions and the additional RNA polymerases needed if mRNAs were unstable. In Fig. 5b) we show that total metabolic cost depends on the average translation number, *p*, reaching a minimum in the range $p=10\text{-}30$. The calculations are detailed in Methods. Obviously this optimum depends on the ribosome initiation frequency and shifts towards more stable mRNAs for lower initiation rates. As have been realized for a long time (21), unstable mRNAs also enable the cell to have a timely and efficient regulation in response to changing growth conditions. Our study shows that unstable mRNAs in addition will be cost efficient because of the saved translation time. These calculations are performed on data from cells growing in glycerol minimal medium. At faster growth rates, the density of ribosomes on the mRNA and the on-rates for initiation is expected to be higher (17) and the advantage of having unstable mRNAs may therefore increase with more rapid growth.

**Discussion**

We have here modelled the translation process in a living bacterial cell and, thanks to the precise input, succeeded in finding a narrow range for the kinetic parameters that reproduces the experimental data. The fortunate experimental finding that one insert did not provoke ribosome queue formation while the tandem insert did, restricted the kinetic parameters. We conclude:

1) The experimental data for translating the *lacZ* mRNA *in vivo* is reproduced by a local model taking into account only stocasticity and variations in individual codon translocation speed, without any need for effects associated to mRNA secondary or tertiary structures. This does not exclude that special mRNA structures have evolved to inhibit the progress of the ribosome in for example the pseudo-knots structures essential for programmed frame-shifting in a wide-spread but rare number of cases (22).

2) The major time-limiting step in translation is stochastic with the broadest possible time distribution, the Poisson distribution. Because the two codons GAA and GAG, read by the same tRNA have different translation rates (2) and because a change in the modification status of wobble-base in the anti-codon loop of this tRNA alters the translation rate dramatically (23), we find it likely that this rate limiting step is bringing the cognate ternary complex into the A-site of the ribosome.

3) Neither a fully deterministic- nor a uniform stochastic mechanism can reproduce the experimental data.

4) In contrast, values spanning ten-fold between the codon specific rates as the tentative values in Table 1 are needed to give the stochastic noise that will give ribosome queue formation on the longest insert, but not on the insert with the same sequence but with half the length.

5) The on-rate of initiating on the *lacZ* ribosome-binding site is close to 1 per sec. Including an occlusion time estimated to 1 sec and 0.2 sec as the time from binding of the 30S ribosome to the formation of the first peptide bond, the resulting initiation frequency on the lacZ mRNA becomes close to 1 per 2.3 sec, in reasonable agreement with the previous determinations (2, 11).

Changing the model parameters (the time for forming the 70S translating ribosome in the interval 0.1 to 0.3 sec, the occlusion length to 10 or 12 codons or decreasing the average number of translations per mRNA to 20 does not change the above conclusions or the finding that unstable mRNAs are more economical to translate. However, minor quantitative changes are seen. If for instance the average number of translations of each mRNA is decreased to 20 the best fitting on-rate will be larger and close to 1.2 per sec.

While our model fulfils the overall feature of increased queuing on the older mRNAs, it does not fully describe the extensive queuing in the case of the long insert in pMAS48GAG. To improve the fit we have explored a number of variations in the model, as well as in the parameters. We have increased the average number of translations per mRNA to 60, varied on-rate rates and changed codon translation rates. In all cases the change affects all 3 constructs proportionally and do not address the difference in queuing for the long compared to the short slow insert. The only way we better can reproduce the additional queuing in long insert, is by assuming that the tRNA$_{Glu}$ becomes rate more limiting for GAG translation in the case of the long insert. This assumption is not in agreement with the finding that rare codons are translated with a rate that is independent of the expression yield (3). However, the effect might be small and not noticed in the previous study or be because the inserts used here use many identical codons. We estimate the use of Glu-tRNA to be increased by approximately 15% under full induction of pMAS48GAG. The drain on glu-tRNA$_{GLU}$ is smaller when inducing pMAS48GAA because the tRNA spends less time on the ribosome in this case. Assuming a varying specific effect for GAG translation, the model can now reproduce the observed queuing difference between pMAS24GAG and pMAS48GAG, provided that we set the on-rate to be at the threshold for queuing.

With respect to the assigned values for the codon specific translation rates given in Table 1, they in addition to reproducing the data of Fig 3a, also reproduce the 35% increased rate of translating the *tuf, tsf, fus* and *rpsA* mRNA with the ribosomal protein type codon usage compared to the *lacI* and *bla* mRNAs, which have a broader codon usage (1). The used A, B and C rates also reproduce the observed translation times of the five additional inserts in (2) and not modelled here.

Our modelling has drawn attention to ribosome collisions and shown them to be frequent and affect the kinetics of translation significantly. Comparing to the measured average rates, especially the fast rates must be corrected upwards due to the effect of collisions because a ribosome on a fast codon is particularly likely to collide with a preceding ribosome when this translate a slow codon. The fastest individual codon translation rate ever measured was the rate of 47 codons per sec for translating codon GAA in the *trmE* strain, deficient in the modification of the anti-codon loop of tRNA$_{GLU}$ (23). Adjusting also this rate upwards as a consequence of collisions, the estimate of the translation K$_{cat}$ value from *in vivo* measurements now gets in total agreement with the K$_{cat}$ value of 50-100 per sec estimated from *in vitro* experiments (24).

Using the codon translation rates from Table 1 to estimate the occlusion times for natural ribosome binding sites of mRNAs, the analysis of 100 such sites indicated that the occlusion time is likely to range from 0.3 to 1.3 sec among different mRNAs. With on-rates of 1 every few seconds, variances in occlusion time will modulate the expression substantially from most if not all ribosome-binding sites.

Finally, our model allows us to obtain accurate estimate on ribosome initiations rates and translation speed of individual codons, which is consistent with a *lacZ* mRNA stability that minimize the overall metabolic cost of producing proteins. We predict that for *lacZ* mRNA, then about 5% of ribosome translation time is wasted because of collisions along the mRNA. Although this waste could be minimized by engineering mRNAs with shorter lifetimes, then the cost of producing more unstable mRNA will exceed the gain associated to increased ribosome efficiency. If one, on the other hand, lower the waste associated to ribosome-ribosome collisions by decreasing ribosome initiation rate this would increase the distance between translating ribosomes. An increase in naked mRNA along the mRNA and also between first ribosome and the RNA-Polymerase may in turn increase the access for degradation factors like Rho, and degradation may be initiated more often at random positions along the mRNA. As 0.4% of all *in-vivo* translations that are terminated prematurely (25) in itself is a large cost in terms of ribosome that waste their time on dysfunctional mRNA, the cost of increasing this termination fraction could easily become larger than the 5% cost of ribosome-ribosome collisions. Overall we therefore find that *E.coli* has optimized both metabolic cost and protein production time by translating each *lacZ* mRNA 30 times with 2.3 sec intervals.

**Methods**

**Description of the translated mRNAs.**

The *lacZ* genes used in this study are derived from the wild type *lacZ* and has the following inserts: pMAS23 has the sequence CGG TCG ACC GAT inserted at codon 927 in *lacZ;* the plasmids pMAS24GAA, pMAS24GAG, pMAS48GAA and pMAS48GAG have the sequences CGA(GAA)$_8$CCG, respectively CGA(GAG)$_8$CCG inserted 3 respectively 6 times in the AccI restriction enzyme site in the CGG TCG ACC GAT sequence in pMAS23 (2).

**Remarks on the experimental method.**

The method determines the time for bringing the radioactivity incorporated into the most N-terminal methionine residue into the full-length protein. The initiating methionine is split of in the case of β-galactosidase, which happens before the protein is completed (27) and the most N-terminal methionine residue is therefore the methionine encoded by codon 3 in *lacZ*. This means that the time the ribosome spends in the initiation process is not included. Similarly, the time the ribosome takes for terminating translation does not influence our measurements, as long as it is not so slow as to provoke queue formation. This is because the bond between the tRNA and the nascent peptide is hydrolysed during our sample preparation that includes boiling with SDS. The full-length position on the gel is therefore achieved before the termination factor has released the peptide. This also means that the time for recycling mRNA, ribosomes, initiation- and termination factors and tRNA does not influence the incorporation kinetics. With respect to equilibrating the radioactivity into the pool of methionine and met-tRNA$_{MET}$ this time has been estimated to be shorter than 5 sec at 25°C i.e. shorter than 2 sec at the 37°C used here (28). This time for pool heating is therefore close to negligible in the 10 sec pulse we use. Furthermore, because we use a mixture of the two cultures, the small time for pool-heating affect the incorporation curves identically for the strains to be compared in an experiment.

**Simulation methods.**

**Stochastic simulation**. We performed Monte Carlo simulations with a constant time step $dt$. We consider a system that consists of multiple mRNAs, which are constantly reshuffled as new mRNAs are transcribed and old ones are decaying. New mRNAs are transcribed stochastically with a rate $R_{birth}$. Ribosomes can bind to existing mRNAs, as long as the binding site is accessible. At each time step, all ribosomes on all mRNAs proceed to the next codon with a probability $R_x dt$ under the condition that the distance to the preceding ribosome is larger than the occluding distance $d$. When a ribosome reaches the 3'end, it terminates the translation with a rate $K_t$. The decay of mRNA with rate $R_{stop}$ is simulated by stopping translation initiation of each mRNA with probability $R_{stop} dt$. For each of these stopped mRNAs, the already initiated ribosomes continue translation until terminating at the 3' end. When all the ribosomes finish the translation, the mRNA is removed from the system.

The dynamics of the model is implemented in timesteps $dt$ which are sufficiently short that even the events with the largest rate occur with probability $<<1$ within $dt$. Each timestep consists of the following actions, taking care of the population of mRNAs and the movement of ribosome on each mRNA:

(a) Add a new mRNA to the system with a probability $R_{birth} dt$.

(b) Perform the following ribosome dynamics (b-1 to b-3) for all the mRNAs in the system:

(b-1) If the first $d$ codons are not occupied by ribosomes a new ribosome can bind to the first codon with probability $K_s dt$. If a new ribosome binds to the first codon the time counter $w$ for that ribosome is set to zero. Also, if the first $d$ codons are not occupied by ribosomes, translation initiation is stopped with a probability $R_{stop} dt$ and the mRNA starts degrading from the 5'end.

(b-2) If there is a ribosome on the codon 1 and $w$ for that ribosome is smaller than the assigned time $\tau=0.2$sec it takes to initiate transcription, then there is no movement. For a ribosome at codon 1 with $w>\tau$, and for any of the other ribosome at positions $x<L$ there is a possibility for movement. For each such ribosome one chose a random number $r$ equally distributed in the interval [0:1], and if $r<R_x dt$ the ribosome is attempting to move one step forward from its current position $x$. This movement, $x \rightarrow x+1$, is performed if and only if there is no other ribosomes in $[x,x+d]$. In case movement is occluded, then the ribosome stays where it is. An eventual ribosome at position $x=L$ cannot move but will finish translation with probability $K_t dt$, and in that case it will leave the mRNA.

(b-3) Add $dt$ to $w$, the duration that a ribosome spent at the first codon.

(c) mRNAs where translation initiation has been stopped and have no ribosomes are removed from the system.

After the procedures (a-c), the time $t$ proceeds by the time step $dt$. Procedure (a-c) is repeated to see the time evolution of the system.

Parameters used in the simulation. The average number of translations per mRNA, $p$, is determined by $R_{birth}/R_{stop}$, which is adjusted such that each mRNA is translated 30 times unless otherwise noted. We use the time step $dt=1/300$ sec for stochastic simulation, while for deterministic simulation $dt$ is set so that $1/(R_x dt)$ becomes an integer to minimize the error in the waiting time. In most of the simulation, we set $R_{birth}=12$, which gives about 2,000 mRNAs in the system at a given time in the steady state when $p=30$.

**Deterministic simulation**. In the deterministic simulations, the rates $R_x$ and $K_t$ in the step (b) are not used to give probabilities to proceed to the next codon but each ribosome waits for the duration $1/R_x$ ($1/K_t$) before it proceeds to the next codon (it terminates the translation). The rest of the simulation has been performed in the same way as the stochastic simulation.

**Reproduction of the incorporation curves.** The incorporation curve is obtained as follows: First, the ensemble of the mRNAs is prepared by performing the simulation for long enough time so that the system reaches the steady state. Then, the time $t$ is reset to zero. From $t=0$ sec to $t=10$ sec, the pulse of radioactive methionine is added. When a ribosome passes a methionine site during this pulse, the ribosome carries one radioactive methionine, and the number of radioactive methionines is added up as the ribosome passes more methionine sites during the pulse. When a ribosome finish the translation, the number of radioactive methionines that the ribosome is carrying is counted as the radioactive signal. The accumulated counts until time $t$, $C(t)$, gives the amount of incorporation, which reaches a constant after long enough time. The $C(t)$ is normalized so that this constant becomes 23, the number of methionine codons on the mRNA. This normalized incorporation curve $C(t)$ are shown in Fig. 3 by dashed lines.

**Metabolic costs**

The metabolic cost estimate is based on cells growing in a minimal medium with doubling time of 60 min and an average mRNA half-life of 1.5 min. Under these conditions a 1 ml culture at a density $OD_{450}$ of 1 (~$4\times10^8$ cells), contain 0.5µg mRNA (~2.2 million nucleotides/cell), 1.1µg of RNAP and 35µg of Ribosomes (17, 26). When calculating metabolic cost in number of ATP per cell, we further use that production of 1g of cellular material requires approximately 2 g of glucose. One glucose molecule weighs 180 Dalton and supplies an energy equivalent of 36 ATP molecules, which each is assumed to be enough for recycling between 1 and 2 nucleic acid bonds in mRNA. First we estimate the mRNA processing cost, including both the mRNA recycling cost and the production cost of the fraction of RNAP that is actively engaged in transcribing mRNA. At a number of proteins per mRNA $p=30$, the cost of recycling the 2.2 million mRNA nucleotides 30 times within one cell generation is between 70 and 140 million ATP per cell. The cost of producing the 10 % of the RNAP that is actively transcribing mRNA (17) is additional 70 million ATP, a number that is uncertain because a fraction of the non-transcribing RNAP may be needed to secure the activity of the active ones. Accordingly we estimate RNAP processing cost to be between 70 and 140 million ATP per cell, which result in a total mRNA processing cost of 140-280 million ATP/cell. When $p$ is decreased, the cost in mRNA processing changes as $1/p$, with the cost=140-280million ATP/cell for $p=30$. Thus the mRNA translation cost increases up to 30 fold as p approaches 1. Ribosomes represent a metabolic cost of 21 billion ATP per wild type cell, which presumably represent a translation speed that is obtained at $p=30$. Compared to p=30, Fig 5a shows that the translation speed change from being 5% faster for $p=1$, to 1.7% slower than for $p>>100$. The corresponding variation in metabolic cost for producing the needed ribosomes therefore change between 20 and 21.3 billion ATP/cell as $p$ is varied from $p=1$ to stable mRNAs. Although the ribosome cost only vary be a few percentages, the fact that ribosomes are expensive in absolute terms makes it possible to save metabolic cost by reducing mRNA stability. This is illustrated in Fig 5b.

**Acknowledgements**

We thank Tom Silhavy, Princeton University for comments on the MS. The Danish National Research Foundation funded this work. NM thanks the Yamada Science Foundation for supporting her stay at the NBI.

**Legends**

**Fig. 1**

Visualization of the modelled translation of *lacZ* mRNA: Ribosomes move from codon to codon on an mRNA, with translation rates that depend on the codon. Each ribosome occupies a space of 11 subsequent codons, and subsequent ribosomes are initiated with a rate given by Ks. The time of forming the 70S initiation complex is usually set to 0.2 sec. The black middle bar indicate the positions of methionine residues in yellow. The lower panel show the distribution of the fast translated A codons (yellow), the middle rate B codons (blue) and the slowly translated C codons (red) in the 1027 codons of the parental *lacZ* gene in pMAS23. When a ribosome reaches the stop codon it is terminated with the non-limiting rate constant of 2 per sec.

**Fig. 2**

Illustration of an idealized experiment where radioactivity is incorporated into full-length protein. The sketch on the ordinate axis indicates that the radioactivity in the early time points originate from the ribosomes close to the 3' end on the mRNA at time of the radioactive pulse.

**Fig. 3.**

a) The experimental data from (2), showing incorporation into β-galactosidase encoded by pMAS23 (+, red), pMAS24GAG (x, green), and pMAS48GAG (*, blue).

b) Best fit to the data. Dotted lines are the simulated curves for the stochastic model and have the same colour code as in panel a.

c) Deterministic simulation with parameters as in b).

d) Stochastic simulation with uniform codon rates of A=B=C=13.25 codons per sec.

For panel b, c and d the data points are as in panel a.

**Fig. 4**

Space-time plot of ribosome traffic as simulated with parameters from figure 3b). Each read dotted line refers to the movement of a single ribosome along the DNA. When ribosomes slow down, for example when queues are developing, one sees increased density of ribosomes. a) Translation of pMAS23, b) pMAS24GAG and c,d) pMAS48GAG. In a,b,c) we translate mRNAs exactly 30 times. In d) we show how a queue develops on a stable mRNA with the long slow insert.

**Fig. 5**

a) Average translation time and b) metabolic cost of translation as function of average number of translations, *p* per mRNA. The metabolic cost is counted by ATP molecules needed to adjust the transcription and translation machinery such that overall protein production is fixed to same value independent of *p* (see Methods). The solid curve is the cost of all ribosomes as function of *p*. The 3 dashed curves represent the total cost, including an mRNA processing cost of 140, 210 and 280 million ATP per cell, respectively, at *p*=30.

**Legend to Tables**

**Table 1.**

The standard genetic code table with our tentative assignment of the codon specific translation rates for *E. coli* growing in glycerol minimal medium. To have only three values for the codon specific translation rates is a simplification and the values we use are tentative, but based on the following reasoning: Four average rates were measured previously: one fast, GAA 22 per sec, two medium GAG 6.4 per sec and CCG 5.8 per sec and one slow, CGA 4.2 per sec (2). From this study one can extrapolate to several likely conclusions: 1) for each amino acid the codon that is predominantly used in the genes encoding ribosomal proteins is assigned the A rate. One exception is the preferred CCG codon that was measured to have a B rate. 2) Other codons translated by the same tRNA are translated three fold slower and is assigned a B rate. The B rate is also assumed for codons that are used with medium frequency in the ribosomal type genes. 3) In some cases as for instance ACT and ACC the frequency of use in the genes for abundant proteins is high and similar, hence both are assigned the A rate. 4) The C rate is assigned to the remaining three non-preferred proline codons, expected to be slower than CCG, and to the rarely used codons ATA, CGA, CGG, AGA, AGG and GGA. The values for the A, B and C rates can be optimized independently in the applet. The rates for translating the codons CCG, GAA, GAG and CGA approximately corrected for collisions are shown in parentheses.

**Table 2:**

Parameter sets that reproduce data almost equally well as the best parameter set: A= 35, B= 8 and C=4.5 codons per sec, p=30 and an on-rate of 0.9 per sec.

**Table 3. Comparison of the conservation of synonymous codons in the 5' and 3' ends of mRNAs.**

The sequences encoding 5 randomly chosen highly expressed genes; two ribosomal proteins, two elongation factors and one RNA polymerase subunit were obtained from the TIGR or Genbank databases. The codons in the 5' end and 3' in each gene, 40 and 60 codons respectively, were compared to the corresponding *E. coli* MG1655 sequence. The table list the percentage of synonymous codon replacements in these gene segments. The database has 1 *Erwinia*, 5 *Escherichia*, 4 *Salmonella*, 6 *Shigella* and 6 *Yersinia* species or variants. The codons used within each of these groups were almost identical and only one example is listed for each family. *E.coli* 536 was the only in the *E. coli* set that differed from MG1655. Not listed are the codon changes to non-synonymous codons.

Figure 1

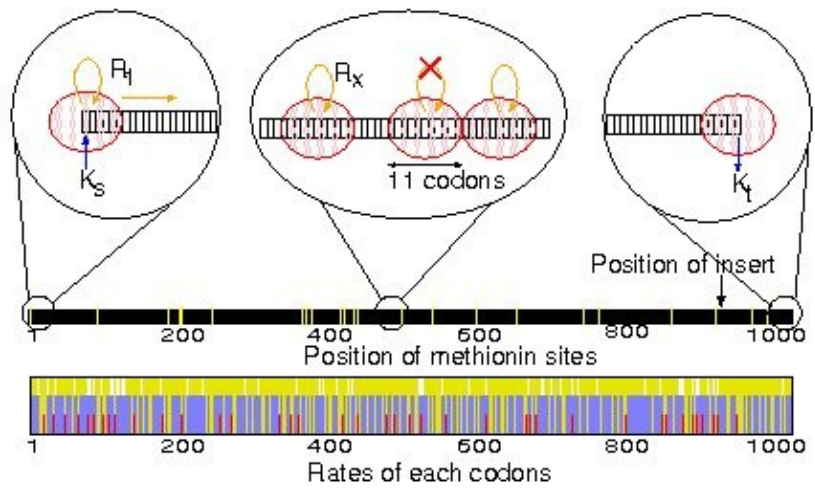

Figure 2

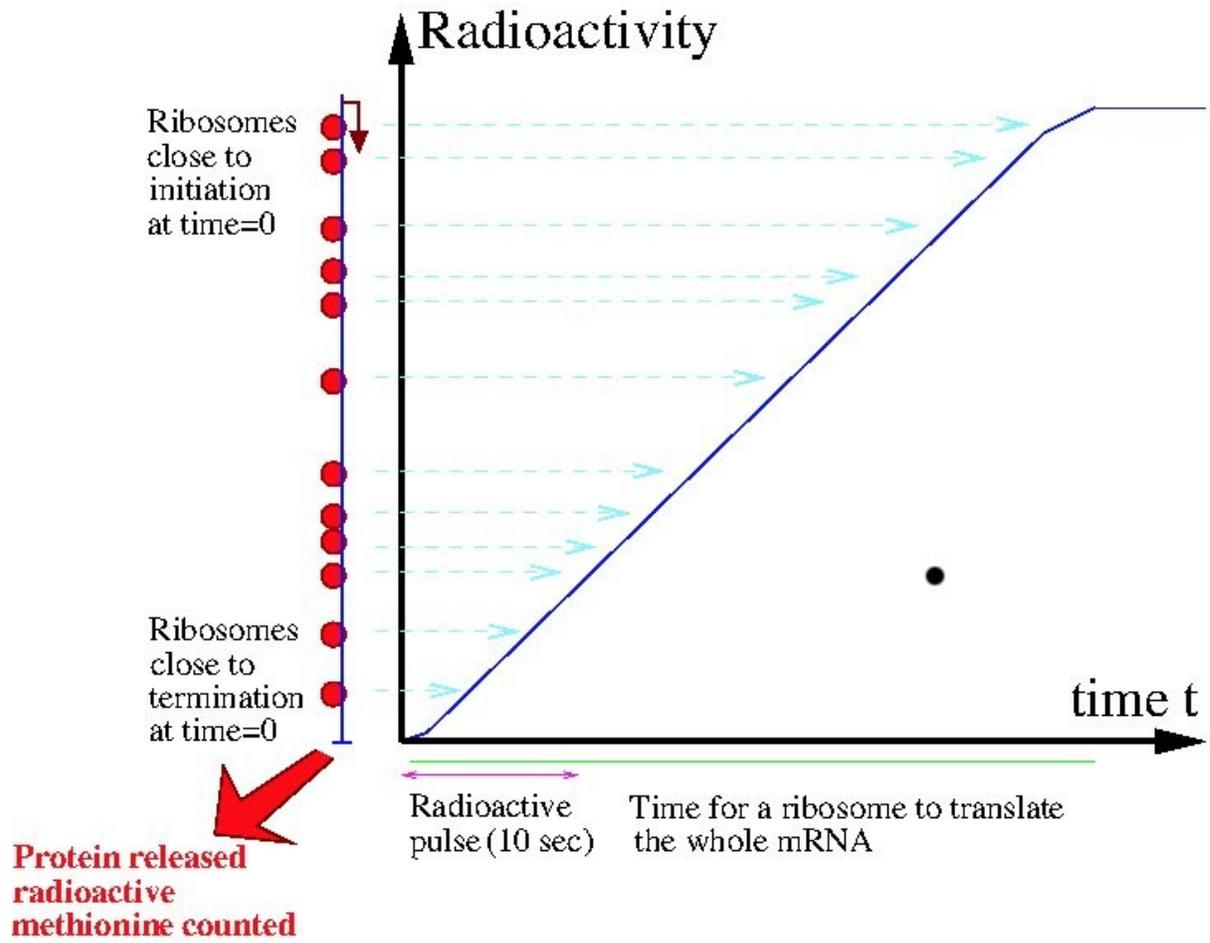

Figure 3

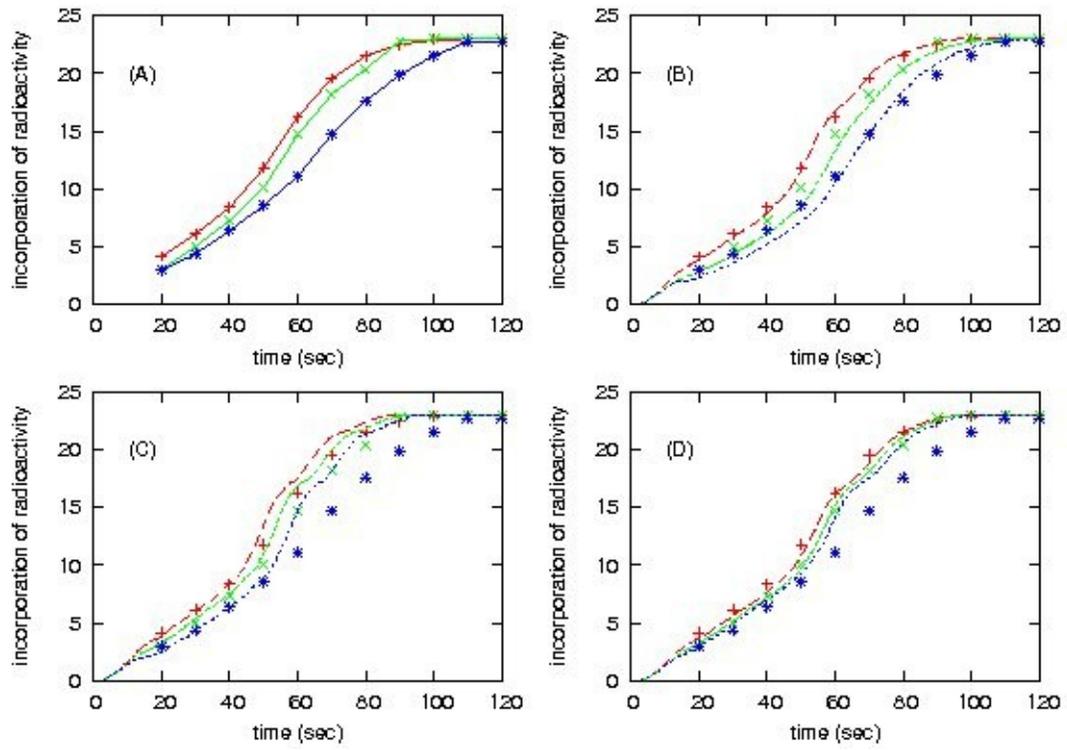

Figure 4

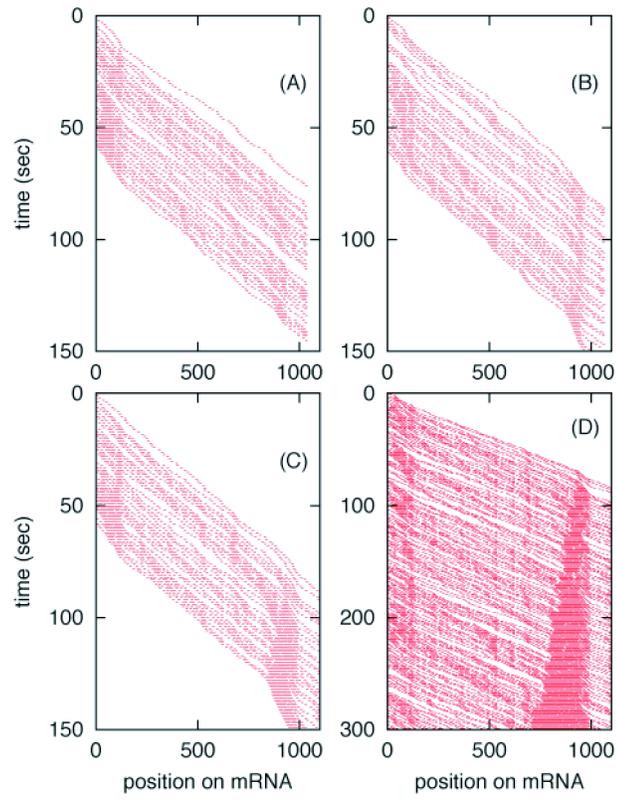

Figure 5

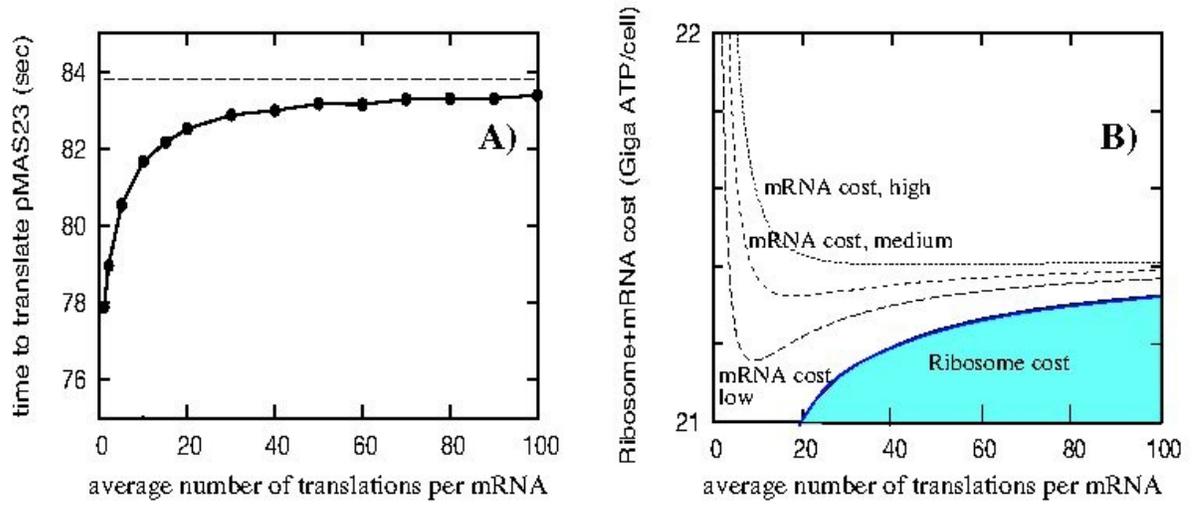

Table 1

| | | | | | | | |
|---|---|---|---|---|---|---|---|
| TTT | B | TCT | A | TAT | B | TGT | B |
| TTC | A | TCC | A | TAC | A | TGC | A |
| TTA | B | TCA | B | TAA | - | TGA | - |
| TTG | B | TCG | B | TAG | - | TGG | A |
| CTT | B | CCT | C | CAT | B | CGT | A |
| CTC | B | CCC | C | CAC | A | CGC | A |
| CTA | B | CCA | C | CAA | B | CGA | C (4.2) |
| CTG | A | CCG | B (7.0) | CAG | A | CGG | C |
| ATT | A | ACT | A | AAT | B | AGT | B |
| ATC | A | ACC | A | AAC | A | AGC | B |
| ATA | C | ACA | B | AAA | A | AGA | C |
| ATG | A | ACG | B | AAG | B | AGG | C |
| GTT | A | GCT | A | GAT | B | GGT | A |
| GTC | B | GCC | B | GAC | A | GGC | A |
| GTA | A | GCA | A | GAA | A (35.0) | GGA | C |
| GTG | A | GCG | B | GAG | B (7.5) | GGG | B |

Table 2

| Rate A (/sec) | Rate B (/sec) | Rate C(/sec) | On rate (/sec) |
|---|---|---|---|
| 35 | 9 | 3.5 | 1.1 to 0.9 |
| 35 | 9 | 3 | 0.8 |
| 35 | 8 | 4.5 | 1.1 to 0.8 |
| 35 | 8 | 4 | 1.0 to 0.8 |
| 30 | 9 | 4 | 1.1 to 1.0 |
| 30 | 9 | 3.5 | 1.0 to 0.8 |

Table 3

| % synonymous codons replaced in segment | rpoC 5' | rpoC 3' | tufB 5' | tufB 3' | tsf 5' | tsf 3' | rpsL 5' | rpsL 3' | rpsG 5' | rpsG 3' |
|---|---|---|---|---|---|---|---|---|---|---|
| E coli 536 | 0 | 2 | 0 | 0 | 0 | 0 | 0 | 4 | 0 | 0 |
| Shig. dysenteria | 0 | 0 | 0 | 0 | 0 | 2 | 0 | 0 | 0 | 0 |
| Salm. paratyphi | 0 | 21 | 7 | 11 | 11 | 23 | 0 | 10 | 3 | 6 |
| Erw. carotovora | 10 | 31 | 5 | 17 | 15 | 23 | 20 | 31 | 15 | 13 |
| Yersinia pestis | 13 | 48 | 26 | 42 | 27 | 29 | 29 | 26 | 26 | 21 |